\documentclass[11pt,onecolumn,twoside]{article}

\usepackage{cite}
\usepackage{amsmath,amssymb,amsfonts}
\usepackage{algorithmic}
\usepackage{graphicx,color}
\usepackage{textcomp}
\usepackage{xcolor}
\usepackage{hyperref}
\hypersetup{hidelinks=true}
\usepackage{algorithm,algorithmic}

\usepackage{url}            
\usepackage{booktabs}       
\usepackage{nicefrac}       
\usepackage{microtype}      
\usepackage{bm}         
\usepackage{cite} 
\usepackage{diagbox}
\usepackage{siunitx}
\usepackage{multirow}
\usepackage{tabularx}
\usepackage{hhline}
\usepackage{arydshln}		
\usepackage{mathtools} 
\usepackage{amsthm}			
\usepackage{setspace}
\usepackage{lipsum} 
\usepackage{adjustbox}
\usepackage{fullpage}

\definecolor{lightgreen}{rgb}{.9,1,.9}

\newcolumntype{L}[1]{>{\raggedright\arraybackslash}p{#1}}
\newcolumntype{C}[1]{>{\centering\arraybackslash}p{#1}}
\newcolumntype{R}[1]{>{\raggedleft\arraybackslash}p{#1}}

\theoremstyle{plain} 
\newtheorem{proposition}{Proposition}

\newtheorem{assumption}{Assumption}

\def\defn{\,\coloneqq\,}

\def\prox{{\mathrm{prox}}}

\def\C{\mathbb{C}}
\def\R{\mathbb{R}}

\def\zerobm{{\bm{0}}}
\def\onebm{{\bm{1}}}

\def\ebm{{\bm{e}}}

\def\xbm{{\bm{x}}}
\def\zbm{{\bm{z}}}
\def\ybm{{\bm{y}}}
\def\rbm{{\bm{r}}}

\def\zbm{{\bm{z}}}

\def\wbm{{\bm{w}}}

\def\Ibm{{\bm{I}}}

\def\thetabm{{\bm{\theta }}}

\def\Ibm{{\bm{I}}}

\def\Ccal{{\mathcal{C}}}

\def\Lcal{{\mathcal{L}}}

\def\Dsf{{\mathsf{D}}}
\def\Gsf{{\mathsf{G}}}

\def\Isf{{\mathsf{I}}}
\def\Rsf{{\mathsf{R}}}
\def\Tsf{{\mathsf{T}}}
\def\Tsf{{\mathsf{T}}}

\def\Dsf{{\mathsf{D}}}

\def\Hsf{{\mathsf{H}}}

\def\Gsf{{\mathsf{G}}}
\def\Isf{{\mathsf{I}}}

\def\Hsf{{\mathsf{H}}}

\def\Bbf{{\mathbf{B}}}

\def\Sbf{{\mathbf{S}}}

\def\Fbf{{\mathbf{F}}}
\def\Hbf{{\mathbf{H}}}

\def\Ibf{{\mathbf{I}}}
\def\Pbf{{\mathbf{P}}}

\def\Hbf{{\mathbf{H}}}
\def\Ibf{{\mathbf{I}}}
\def\Abf{{\mathbf{A}}}

\def\Jbf{{\mathbf{J}}}
\def\Mbf{{\mathbf{M}}}

\def\xbmast{{\bm{x}^\ast}}
\def\xbmbar{{\overline{\bm{x}}}}

\def\xbmhat{{\widehat{\bm{x}}}}

\def\argmin{\mathop{\mathrm{arg\,min}}} 

\def\BibTeX{{\rm B\kern-.05em{\sc i\kern-.025em b}\kern-.08em
    T\kern-.1667em\lower.7ex\hbox{E}\kern-.125emX}}
\AtBeginDocument{\definecolor{tmlcncolor}{cmyk}{0.93,0.59,0.15,0.02}\definecolor{NavyBlue}{RGB}{0,86,125}}

\title{Deep Equilibrium Learning of Explicit Regularizers \\ for Imaging Inverse Problems}

\author{Zihao~Zou
\thanks{The first two authors contribute equally.}
\thanks{Department of Computer Science \& Engineering, Washington University in St.\ Louis, St.\ Louis, MO 63130}
\hspace{0.05em},
Jiaming~Liu$^{\ast}$%
\thanks{Department of Electrical \& Systems Engineering, Washington University in St.~Louis, St.~Louis, MO 63130.}
\hspace{0.05em}, 
Brendt~Wohlberg%
\thanks{Theoretical Division, Los Alamos National Laboratory, Los Alamos, NM 87545 USA.},
\hspace{0.05em} \\
and Ulugbek~S.~Kamilov$^{\dagger, \ddagger}$%
}

\begin{document}
\date{}
\maketitle

\begin{abstract}
There has been significant recent interest in the use of deep learning for regularizing imaging inverse problems. Most work in the area has focused on regularization imposed \emph{implicitly} by convolutional neural networks (CNNs) pre-trained for image reconstruction. In this work, we follow an alternative line of work based on learning \emph{explicit} regularization functionals that promote preferred solutions. We develop the \emph{Explicit Learned Deep Equilibrium Regularizer (ELDER)} method for learning explicit regularizers that minimize a mean-squared error (MSE) metric. ELDER is based on a regularization functional parameterized by a CNN and a deep equilibrium learning (DEQ) method for training the functional to be MSE-optimal at the fixed points of the reconstruction algorithm. The explicit regularizer enables ELDER to directly inherit fundamental convergence results from  optimization theory. On the other hand, DEQ training enables ELDER to improve over  existing explicit regularizers without prohibitive memory complexity during training. We use ELDER to train several approaches to parameterizing explicit regularizers and test their performance on three distinct imaging inverse problems.
Our results show that ELDER can greatly improve the quality of explicit regularizers compared to existing methods, and show that learning explicit regularizers does not compromise performance relative to methods based on implicit regularization.
\end{abstract}

\section{Introduction}
{T}{he} recovery of an unknown image from a set of noisy measurements is one of the most widely-studied problems in computational imaging. The task is often formulated as an \emph{inverse problem}, and solved by integrating the measurement model characterizing the response of the imaging instrument with a regularizer imposing prior knowledge of the unknown image. Over the years, many regularizers have been proposed as image priors, including those based on transform-domain sparsity, low-rank penalty, and self-similarity. The focus in the area has recently shifted to methods based on deep learning (DL)~\cite{McCann.etal2017, Lucas.etal2018, Gilton.etal2020}. Instead of explicitly defining a regularizer, DL approaches for solving inverse problems learn a mapping from the measurements to the desired image by training a convolutional neural network (CNN) to perform regularized inversion.

Model-based DL (MBDL) has emerged as an alternative to the traditional DL, where the knowledge of the measurement model is combined with an image prior specified by a CNN (see reviews~\cite{Ongie.etal2020, Kamilov.etal2023}). For example, plug-and-play priors (PnP) is a widely-used MBDL framework based on specifying the image prior using a pre-trained image denoiser~\cite{Venkatakrishnan.etal2013, Sreehari.etal2016, Kamilov.etal2023}. Deep unfolding (DU) is another MBDL approach based on interpreting a fixed-number of iterations of an image recovery algorithm as layers of a neural network trained end-to-end in a supervised fashion~\cite{zhang2018ista, Hauptmann.etal2018}. DU architectures, however, are usually limited to a small number of unfolded iterations due to the high memory complexity of training. Deep equilibrium learning (DEQ) is an alternative to DU that enables training of very deep neural networks with a constant memory complexity in the number of iterations~\cite{Bai.etal2019, Gilton.etal2021, Fung.etal2021}. 

Existing research on MBDL for inverse problems, including most of the work on PnP, DU, and DEQ, has largely focused on regularization \emph{implicit} in pre-trained neural networks. While this strategy has led to state-of-the-art performance, it requires stringent assumptions on the MBDL architecture to ensure algorithmic stability (see related reviews~\cite{Kamilov.etal2023, Mukherjee.etal2023}). For example, a common assumption used for proving stability of MBDL architectures is that the CNN is nonexpansive~\cite{Ryu.etal2019, Sun2019b, Gilton.etal2021, Pramanik.Jacob2022, Liu.etal2022a}. An alternative line of work has explored MBDL with learned \emph{explicit} regularizers parameterized by neural networks~\cite{Romano.etal2017, Cohen.etal2021, Hurault.etal2022}. Explicit regularization functionals significantly simplify the convergence analysis due to the direct applicability of optimization theory.

We develop a new method called \emph{Explicit Learned Deep Equilibrium Regularizer (ELDER)} that improves over existing approaches for learning explicit regularization functionals. This method seeks to learn a regularizer that achieves the smallest value of mean-squared error (MSE) for a given inverse problem. To this end, we parameterize the regularization functional by a CNN and train it end-to-end by using DEQ on the MSE loss. Similarly to existing approaches for learning explicit regularizers, ELDER directly inherits traditional concepts from optimization for parameter selection and convergence analysis. On the other hand, ELDER outperforms existing explicit regularization methods due to its ability to optimize learned functionals at the fixed points of the reconstruction algorithm. We present numerical results on three imaging inverse problems: image super-resolution, image reconstruction from subsampled Fourier measurements, and image inpainting. We apply ELDER to optimize the weights of three parameterization approaches for regularization functionals---namely least squares residual (LSR), regularization by denoising (RED), and direct scalar-valued network (DSV)---and compare the effectiveness of all three as regularizers when trained to be MSE-optimal. We also show that ELDER does not compromise the imaging quality relative to methods based on implicit regularization. Our results show that ELDER achieves excellent imaging results, while also offering the potential for automatic step-size selection and algorithmic stability, even when using expansive update rules. In short, our work provides a method to learn state-of-the-art explicit regularizers for MBDL that preserve traditional tools from optimization theory~\footnote{Our code is publicly available at \url{https://github.com/wustl-cig/ELDER}}.

\section{Background}

\textbf{Inverse Problems.} We consider the imaging inverse problem of recovering an unknown image $\xbm \in \R^n$ from noisy measurements $\ybm = \Abf\xbm + \ebm,$ where $\Abf$ is the \emph{measurement operator} and $\ebm\in\R^m$ is additive white Gaussian noise (AWGN) vector. The problem is traditionally formulated as an optimization problem 
\begin{equation}
\label{Eq:OptimizationForInverseProblem}
\xbmhat = \argmin_{\xbm \in \R^n} f(\xbm) \quad\text{with}\quad f(\xbm) = g(\xbm) + \tau h(\xbm),
\end{equation}
where $\tau > 0$ is the regularization parameter, $g$ is the \emph{data-fidelity term} enforcing consistency of the solution with $\ybm$, and $h$ is the \emph{regularizer} enforcing prior knowledge of $\xbm$. 

\medskip\noindent
\textbf{Model-based Optimization.} Proximal algorithms are often used for solving problems of form~\eqref{Eq:OptimizationForInverseProblem} when $h$ is nonsmooth (see the review~\cite{Parikh.Boyd2014}). One widely used family of proximal algorithms for imaging inverse problems are the proximal gradient method (PGM)~\cite{Figueiredo.Nowak2003}. PGM avoids differentiating $h$ by using the \emph{proximal operator}, which can be defined as
\begin{equation}
\label{eq:proximaloperator}
\prox_{\tau h} (\zbm) := \argmin_{\xbm \in \R^n} \bigg\{ \frac{1}{2} \|\xbm -\zbm\|^2_2 + \tau h(\xbm) \bigg\}, 
\end{equation}
with $\tau > 0$, for any proper, closed, and convex function $h$~\cite{Parikh.Boyd2014}. Comparing~\eqref{eq:proximaloperator} and~\eqref{Eq:OptimizationForInverseProblem}, we see that the proximal operator can be interpreted as a MAP estimator for the problem
\begin{equation}
\label{Eq:MAPDenoise}
\zbm = \xbm_0 + \wbm \quad\text{where}\quad \xbm_0 \sim p_{\xbm_0}\,, \quad \wbm \sim \mathcal{N}(0, \tau \Ibm)\,,
\end{equation}
by setting $h(\xbm) = -\log(p_{\xbm_0}(\xbm))$. It is worth noting that another less well-known but equally valid statistical interpretation of the proximal operator is as a minimum mean-squared error (MMSE) estimator~\cite{Gribonval2011, Xu.etal2020}.

\medskip\noindent
\textbf{PnP.} PnP refers to a family of algorithms that integrate measurement operators and CNN denoisers for solving inverse problems (see the recent review~\cite{Kamilov.etal2023}). Since the prior in PnP is implicit in the denoiser, it is common to interpret PnP as fixed-point iterations of some high-dimensional operators~\cite{Reehorst.Schniter2019}. For example, given a denoiser $\Dsf_\thetabm: \R^n \rightarrow \R^n$ parameterized by a CNN with weights $\thetabm$, the iterations of proximal gradient method (PGM)~\cite{Figueiredo.Nowak2003} variant of PnP can be written
\begin{align}
\label{Eq:PnPPGM}
\xbm^k = \Tsf_\thetabm(\xbm^{k-1})\;\;\text{with}\;\;\Tsf_\thetabm\defn\Dsf_\thetabm(\Isf - \gamma \nabla g)\;,
\end{align}
where $g$ is the data-fidelity term in~\eqref{Eq:OptimizationForInverseProblem}, $\Isf$ is the identity mapping, and  $\gamma > 0$ is the step size. 

\medskip\noindent
\textbf{DU and DEQ.} DU is a DL paradigm that has gained popularity due to its ability to systematically connect iterative algorithms and deep neural network architectures (see reviews in~\cite{Ongie.etal2020, Monga.etal2021}). DEQ~\cite{Bai.etal2019} is a related approach that enables training of infinite-depth, weight-tied networks by analytically backpropagating through the fixed points using implicit differentiation. The DEQ output is specified implicitly as a fixed point of an operator $\Tsf_{\thetabm}$ parameterized by weights $\thetabm$ 
\begin{equation}
\label{Eq:DEQFixedPoint}
\xbmbar = \Tsf_{\thetabm}(\xbmbar)\; .
\end{equation}
The DEQ forward pass usually estimates $\xbmbar$ in~\eqref{Eq:DEQFixedPoint} by running a fixed-point iteration. The comparison of equations~\eqref{Eq:PnPPGM} and~\eqref{Eq:DEQFixedPoint} highlights the connection between PnP and DEQ, which was recently explored~\cite{Gilton.etal2021} by using DEQ for end-to-end learning of the weights of the CNN prior $\Dsf_\thetabm$. There has been considerable interest in DEQ for imaging, including in MRI~\cite{Gilton.etal2021}, computed tomography (CT)~\cite{Liu.etal2022a} and video snapshot imaging~\cite{Zhao.etal2022}.

The convergence of forward iterations is essential for the stability and accuracy of DEQ. Similar to the theoretical analysis of PnP~\cite{Ryu.etal2019, Combettes.2021etal}, a sufficient condition to guarantee the convergence of the DEQ forward pass is to ensure that the residual $\Rsf_\thetabm\defn\Isf-\Dsf_\thetabm$ of $\Dsf_\thetabm$ is \emph{Lipschitz continuous} with constant $0<\alpha\leq 1$~\cite{Gilton.etal2021}
\begin{equation}
\label{Eq:Lipschitz}
\|\Rsf_\thetabm(\xbm) - \Rsf_\thetabm(\zbm)\|_2 \leq \alpha\|\xbm - \zbm\|_2,\quad\forall \xbm,\zbm\in\R^n.
\end{equation}
Since most CNN architectures do not inherently satisfy this property, several methods have been proposed to train Lipschitz constrained CNNs~\cite{Miyato.etal2018, Ryu.etal2019}. However, it has been observed that constraining CNNs to be Lipschitz continuous can negatively impact their performance~\cite{Hurault.etal2022, Goujon.etal2022}.

\medskip\noindent
\textbf{Explicit Regularizers.}
RED~\cite{Romano.etal2017} is an early approach for specifying explicit regularizers for inverse problems by parameterizing them using an image denoiser $\Dsf_\thetabm$ 
\begin{align}
\label{Eq:RED-ED}
h(\xbm)=\frac{1}{2}\xbm^{\Tsf}(\xbm - \Dsf_\thetabm(\xbm)).
\end{align}
When the denoiser is locally homogeneous and has a symmetric Jacobian~\cite{Romano.etal2017, Reehorst.Schniter2019}, the gradient of the RED regularizer $h$ has a simple expression 
\begin{equation}
\label{Eq:RED-Grad}
\nabla h(\xbm)=\xbm-\Dsf_{\thetabm}(\xbm),
\end{equation}
which enables efficient minimization of the RED functional within~\eqref{Eq:OptimizationForInverseProblem}. However, when these rather stringent conditions are not satisfied, RED does not correspond to an explicit regularizer, corresponding instead to an MBDL method with an implicit regularizer~\cite{Reehorst.Schniter2019}. 


Other notable approaches for explicit regularization include adversarial regularization~\cite{Lunz.etal2018} and its convex counterpart~\cite{Mukherjee.etal2020}, network Tikhonov (NETT)~\cite{Li.etal2020}, and total deep variation (TDV)~\cite{Kobler.etal2020}. All these approaches seek to explicitly parametrize the regularization functional using a neural network. Another line of work has explored gradient-step denoisers for PnP based on the direct parameterization of regularization functionals, thus leading to explicit loss functions and convergence guarantees without Lipschitz constraints on the neural networks~\cite{Cohen.etal2021, Hurault.etal2022, Fermanian.etal2022}. 

\medskip\noindent
\textbf{Our contributions.} We propose a novel approach for learning explicit regularizers $h$ for MBDL by directly parameterizing the regularizer with a CNN and training it to be end-to-end optimal using DEQ. Our method inherits all the benefits of having an explicit regularizer, such as convergence without Lipschitz constraints on the CNN and the possibility of using line-search strategies for step-size selection. At the same time, our method leads to state-of-the-art explicit regularizers by being trained to be MSE optimal at the fixed-point of the inference algorithm. Our results show that explicit regularizers trained end-to-end can match the performance on implicit regularization obtained via traditional DEQ. 

\section{Proposed Method}

We now present our proposed method for learning potential functions. Unlike the traditional DEQ approach for inverse problems~\cite{Gilton.etal2021}, the forward pass in our method is designed to minimize an explicit objective, where the regularization functional is parameterized by a CNN. 


\subsection{ELDER Forward Pass}

We consider an inverse problem with a regularization function $h_\thetabm:\R^n\rightarrow\R$ parameterized by weights $\thetabm$
\begin{equation} 
\label{Eq:OptimizationForInverseProblem2}
\xbmhat = \argmin_{\xbm \in \R^n} f_\thetabm(\xbm) \quad\text{with}\quad f_\thetabm(\xbm) = g(\xbm) + \tau h_\thetabm(\xbm).
\end{equation}
The forward pass of ELDER seeks to minimize $f_\thetabm$ via the iterations
\begin{align}
\label{Eq:fwddeq}
\xbm^k = \Tsf_\thetabm(\xbm^{k-1})\;\;\text{with}\;\;\Tsf_\thetabm\defn\prox_{\gamma g} (\Isf - \gamma\tau\nabla h_\thetabm)\;,
\end{align}
where $\gamma > 0 $ is the step-size. For a linear inverse problems with a $\ell_2$-norm loss $g(\xbm)=\frac{1}{2}\|\ybm - \Abf\xbm\|_2^2$, the proximal operator has a closed-form solution
\begin{equation}
\label{Eq:Prox_GS-PnP}
\prox_{\gamma g} (\zbm) = (\Ibf+\gamma \Abf^\Hsf\Abf)^{-1} (\zbm + \gamma \Abf^\Hsf \ybm),
\end{equation}
where $\Ibf$ is the identity matrix and $\Abf^{\Hsf}$ denotes the conjugate transpose of $\Abf$. In many applications, the proximal map~\eqref{Eq:Prox_GS-PnP} can be computed or approximated efficiently using general methods---such as conjugate gradient---or with specialized methods---such as when the forward model is a spatial blurring operator that can be computed using the fast Fourier transform (FFT)~\cite{Teodoro.etal2019, Zhang.etal2021b, Kamilov.etal2023}.

The step-size $\gamma$ must be carefully selected to ensure the convergence of the algorithm in eq.~\eqref{Eq:fwddeq}. Since we have access to the explicit objective function, we can ensure convergence by adopting a line-search strategy. To that end, we use a backtracking line-search (BLS) to ensure the sufficient decrease condition (see Section 9.2 in~\cite{Boyd.Vandenberghe2004})
\begin{equation}
\label{Eq:Backtrack}
\begin{aligned}
f(\xbm^{k-1}) - f(\xbm^{k}) \geq \frac{\rho}{\gamma}\|\xbm^{k}-\xbm^{k-1}\|_2^2, \quad 0 < \rho < \frac{1}{2}.
\end{aligned}
\end{equation}
Our implementation starts with a large step-size $\gamma>0$, which is subsequently reduced by a factor $0 < \beta <1$ to ensure that~\eqref{Eq:Backtrack} is satisfied.


\subsection{Parameterization of Regularization Functionals}
\label{Sec:ConstructPrior}

\noindent
Several approaches for explicitly parametrizing regularization functionals using CNNs have been previously proposed. We consider three well-known approaches and compare them in our numerical evaluations. All three regularizers are based on an operator $\Gsf_\thetabm:\R^n\rightarrow\R^n$ corresponding to an image-to-image CNN. Note that $\Gsf_\thetabm$ can be implemented using any differentiable CNN architecture.

\medskip\noindent
\textbf{Least-Squares Residual (LSR).} Inspired by one of the formulations suggested in~\cite{Romano.etal2017} (see Section 5.2) and in the recent work on the gradient-step denoiser for PnP~\cite{Hurault.etal2022}, we first consider a regularizer that explicitly quantifies the distance between the input and the output of $\Gsf_\thetabm$. More specifically, we consider
\begin{equation}
\begin{aligned}
    \label{Eq:regularizer1}
    &h_\thetabm^{\textsf{\tiny LSR}}(\xbm)=\frac{1}{2}\|\xbm-\Gsf_\thetabm(\xbm)\|_2^2\quad\text{and}\\
    &\nabla h_\thetabm^{\textsf{\tiny LSR}}(\xbm) = (\Ibf - \Jbf_{\Gsf_\thetabm}(\xbm))^{\Tsf}(\xbm-\Gsf_\thetabm(\xbm)),
\end{aligned}    
\end{equation}
where $\Jbf_{\Gsf_\thetabm}$ is the Jacobian of $\Gsf_\thetabm$ with respect to $\xbm$. LSR can be interpreted as promoting solutions to the inverse problem that are near the fixed points of the CNN $\Gsf_\thetabm$.  

\medskip\noindent
\textbf{(Real) RED Regularizer.} 

While the RED gradient in~\eqref{Eq:RED-Grad} has been extensively used for solving inverse problems, the RED functional~\eqref{Eq:RED-ED} has not been previously used as a regularizer. We explore the RED functional itself as a regularizer by considering the true gradient of $h_\thetabm^{\textsf{\tiny RED}}$ given by
\begin{equation}
\begin{aligned}
    \label{Eq:regularizer2}
    \nabla h^{\textsf{\tiny RED}}_\thetabm(\xbm) = \xbm -\frac{1}{2}\Gsf_\thetabm(\xbm) - \frac{1}{2}[\Jbf_{\Gsf_\thetabm}(\xbm)]^\Tsf\xbm,
\end{aligned}    
\end{equation}
where we do not require that the network $\Gsf_\thetabm$ is locally homogeneous and has a symmetric Jacobian.

\medskip\noindent
\textbf{Direct Scalar-Valued (DSV) Regularizer.} Similar to~\cite{Cohen.etal2021} and~\cite{Mukherjee.etal2021}, we can also directly sum the output of the CNN $\Gsf_\thetabm$ to obtain a scalar-valued neural network
\begin{equation}
h_\thetabm^{\textsf{\tiny DSV}}(\xbm)= \onebm^\Tsf\Gsf_\thetabm(\xbm)\quad{\text{and}}\quad\nabla h_\thetabm^{\textsf{\tiny DSV}}(\xbm) = [\Jbf_{\Gsf_{\thetabm}}(\xbm)]^{\Tsf}\onebm.
\end{equation}


\subsection{Jacobian-free Deep Equilibrium Learning}
\label{Sec:TrainingPriors}

We train the regularizer $h_\thetabm$ by minimizing the discrepancy between a fixed-point $\xbmbar = \Tsf_{\thetabm}(\xbmbar)$ obtained via~\eqref{Eq:fwddeq} and the ground-truth image $\xbm$ using MSE loss $\Lcal(\thetabm)= \frac{1}{2}\|\xbmbar(\thetabm)-\xbm\|_2^2$.
The DEQ backward pass produces gradients by implicitly differentiating through the fixed points
\begin{equation}
\label{Eq:DEQexact}
\begin{aligned}
\nabla \Lcal(\thetabm) = \left[\Jbf_{\Tsf_\xbmbar}(\thetabm)\right]^{\Tsf}\left(\Ibf - \Jbf_{\Tsf_\thetabm}(\xbmbar)\right)^{-\Tsf}(\xbmbar-\xbm).
\end{aligned}
\end{equation}
This converts the memory-intensive task of backpropagating through many iterations of $\Tsf_\thetabm$ to the problem of calculating an inverse Jacobian-vector product. Since inverting the Jacobian matrix in~\eqref{Eq:DEQexact} can become computationally expensive, we introduce an approximation that replaces the inverse-Jacobian term in~\eqref{Eq:DEQexact} with an identity as in ~\cite{Fung.etal2021, Geng2021.etal}
\begin{equation}
\label{Eq:DEQinexact}
\begin{aligned}
\nabla \Lcal(\thetabm)\approx\nabla \Lcal_{\textsf{\tiny JFB}}(\thetabm)= \left[\Jbf_{\Tsf_\xbmbar}(\thetabm)\right]^{\Tsf}(\xbmbar-\xbm).
\end{aligned}
\end{equation}
This \emph{Jacobian-free approximation} significantly reduces the complexity of the backward pass without compromising the quality of DEQ.

\subsection{Convergence Theory}
\label{Sec:ANALYSIS}

Since ELDER minimizes an explicit loss $f_\thetabm = g + \tau h_\thetabm$, it directly inherits traditional convergence results from optimization theory~\cite{Attouch.etal2013}. For completeness, we state these convergence results using our notation.


\begin{assumption}
\label{As:Assum1}
The function $g$ is proper, convex, and lower semi-continuous. The function $h_\thetabm:\R^n\rightarrow\R$ is proper, lower semi-continuous, finite valued, and differentiable with L-Lipschitz gradient.
\end{assumption}
These assumptions on $g$ and $h_\thetabm$ are standard, and are satisfied by a large number of functions used in the context of inverse problems (see, for example, the discussion in~\cite{Hurault.etal2022}).

\begin{proposition}
\label{Thm:Thm1}
Run the PGM in~\eqref{Eq:fwddeq} under Assumption~\ref{As:Assum1} using $\gamma \in (0, 1/(\tau L)]$. Then, the sequence $\{f_\thetabm(\xbm^t)\}_{t \geq 1}$ monotonically decreases and $\nabla f_\thetabm(\xbm^t) \rightarrow \zerobm$ as $t \rightarrow \infty$.
\end{proposition}
The proof is widely available in the existing analyses of PGM~\cite{Xu.etal2020, Hurault.etal2022}. The importance of this result is that, unlike the traditional DEQ based on implicit regularization~\cite{Gilton.etal2021}, the stability of the ELDER forward pass does not require any assumptions on the non-expansiveness of the CNN prior. 


\subsection{Additional Technical Details} 
\label{Sec:Implementation}
Our method is compatible with any differentiable CNN architecture for implementing $\Gsf_\thetabm$. We use the simplified DRUNet architecture~\cite{Zhang.etal2021b} for ELDER and the traditional DEQ~\cite{Gilton.etal2021}. We have replaced the rectified linear unit (ReLU) activations with exponential linear unit (ELU) ones to ensure the smoothness of $h_\thetabm$. We also limit the number of residual blocks at each scale to 2. Similar to~\cite{Gilton.etal2021}, the CNN prior of ELDER is initialized using a pre-trained denoisers. Additionally, we follow ~\cite{Gilton.etal2021, bai.etal2022} in setting the convergence criterion in training to
\begin{equation}
\|\xbm^{k} - \xbm^{k-1}\|_2/\|\xbm^{k-1}\|_2 < \epsilon\,,
\end{equation}
where we choose $\epsilon=10^{-2}$ for all experiments.
\begin{table*}[h!]
\centering
\caption{Average PSNR (dB) results of different methods for noise levels 5, 15, 25 and 50 on BSD68, CBSD68 and Kodak24 datasets.
The \textbf{best} and \underline{second best} results are highlighted, respectively}
\label{Tab:Denoising}
\resizebox{0.6\textwidth}{!}{%
\begin{tabular}{lccccccccc}
\hline
\multirow{2}{*}{Datasets} && \multirow{2}{*}{Denoiser} &&& \multicolumn{4}{c}{Noise Level} \\ 
       \cline{6-9}       &&              &&            & 5      & 15     & 25    & 50  &   \multirow{-2}{*}{Average}  \\ \hline\hline
\multirow{4}{*}{BSD68} && DRUNet     &&              & \bf38.04  & \bf31.89  & \bf29.43 & \bf26.56 & 31.48 \\
                       && LSR       &&              & \underline{37.98}  & \underline{31.85}  & \underline{29.42} & \underline{26.52} & 31.44 \\
                       && RED       &&              & 37.83  & 31.77  & 29.36 & 26.48 & 31.36\\
                       && DSV      &&              & 37.63  & 31.52  & 29.10  & 26.18 & 31.11\\ \hline
\multirow{4}{*}{CBSD68} && DRUNet    &&              & \bf40.64  & \bf34.31  & \bf31.69 & \bf28.49 & 33.78 \\
                       && LSR      &&               & \underline{40.55}  & \underline{34.24}  & \underline{31.62} & \underline{28.42} & 33.71 \\
                       && RED      &&               & 40.23  & 34.06  & 31.46 & 28.27 & 33.64 \\
                       && DSV      &&               & 40.24  & 34.05  & 31.45 & 28.25 & 33.50\\ \hline
\multirow{4}{*}{Kodak24} && DRUNet    &&              & \bf40.95  & \bf35.32  & \bf32.90 & \bf29.87 & 34.76\\
                       && LSR      &&               & \underline{40.78}  & \underline{35.19}  & \underline{32.77} & \underline{29.71} & 34.61 \\
                       && RED      &&               & 40.33  & 34.95  & 32.54 & 29.47 & 34.32\\
                       && DSV      &&               & 40.43  & 34.92  & 32.53 & 29.46 & 34.34\\ \hline
\end{tabular}%
}
\end{table*}


\section{Numerical Evaluation}

\subsection{Comparing Parametrization Strategies}

\label{Sec:Imagedenoising}

We first compare the performance of the three parameterization strategies in Section~\ref{Sec:ConstructPrior} on image denoising. We pre-train all the regularizers by adopting the gradient-step denoising strategy from~\cite{Hurault.etal2022}. Pre-training explicit regularizers as denoisers is computationally useful for ELDER, since the denoisers can be used to initialize the DEQ learning. All three explicit denoisers are compared against DRUNet~\cite{Zhang.etal2021b}, which was shown to result in state-of-the-art PnP image restoration. 

We employ the color image training dataset in~\cite{Zhang.etal2021b}, which is a combination of the 400 CBSD images, 4,744 images of Waterloo Exploration Database~\cite{Ma.etal2016}, 900 images from DIV2K dataset, and 2,750 images from Flickr2K dataset. During training we consider AWGN with standard deviation $\sigma$ at a uniform random draw from the range $[0, 55/255]$.
\begin{figure*}[t!]
	\centering
	\includegraphics[width=0.99\textwidth]{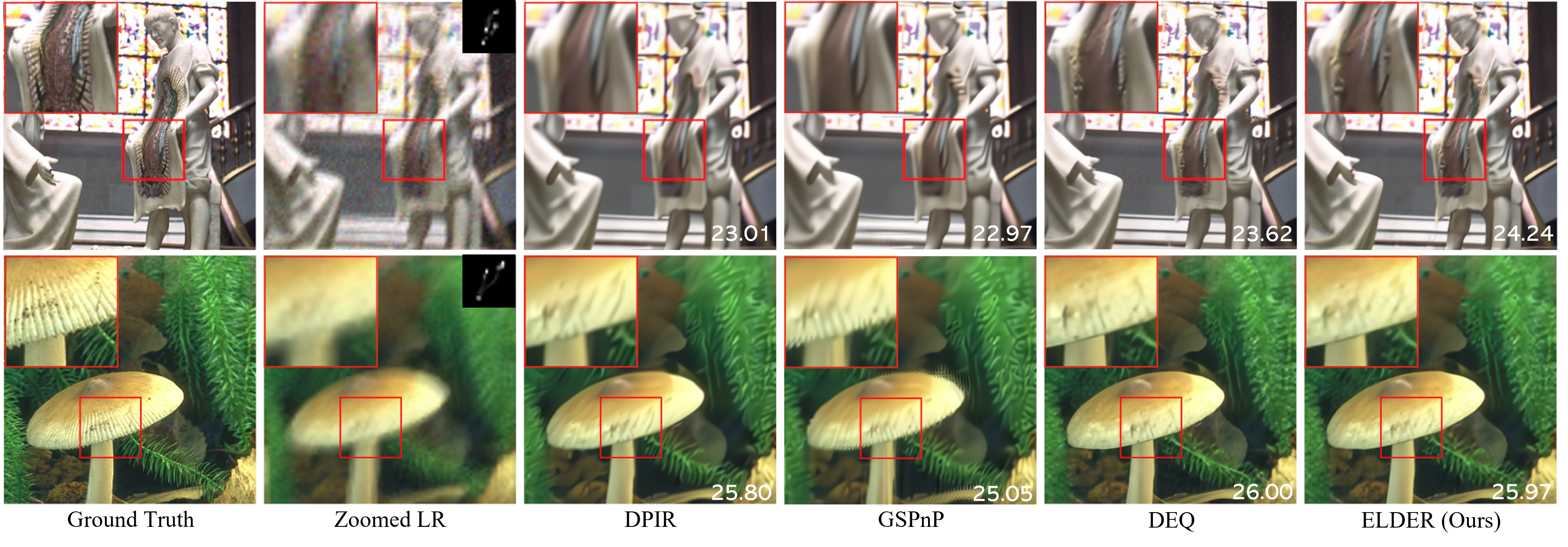}
	\caption{Visual comparison of ELDER against several well-known methods on the problem of single image super-resolution (SISR) with scale $d=3$ (top) and $d=4$ (bottom). The blur kernels are shown in the zoomed low-resolution (LR) images, respectively. Note how ELDER significantly improves over GSPnP, a recent method for learning explicit regularizers. Additionally, it performs better than DEQ and DPIR, both of which rely on implicit regularization specified by a CNN. This figure shows that ELDER learns explicit regularizers without compromising  image quality.}
	\label{fig:sr_visual}
\end{figure*}

Table~\ref{Tab:Denoising} presents the results of all denoisers at several noise levels ($\sigma\in\{5,15,25,50\}$) on both color and gray images. The RED denoisier in the table refers to the gradient-step in~\eqref{Eq:regularizer2}. While all three explicit denoisers perform well at all noise levels, LSR most closely matches the performance of DRUNet. It is worth noting that LSR (with 2 residual blocks at each scale) uses fewer parameters than DRUNet (with 4 residual blocks). Note that the PSNR gap between DRUNet and LSR is within 0.15 dB, which implies that having explicit regularizers does not significantly impair performance (this was also observed in~\cite{Hurault.etal2022}). 

\begin{figure*}[t!]
	\centering
	\includegraphics[width=0.99\textwidth]{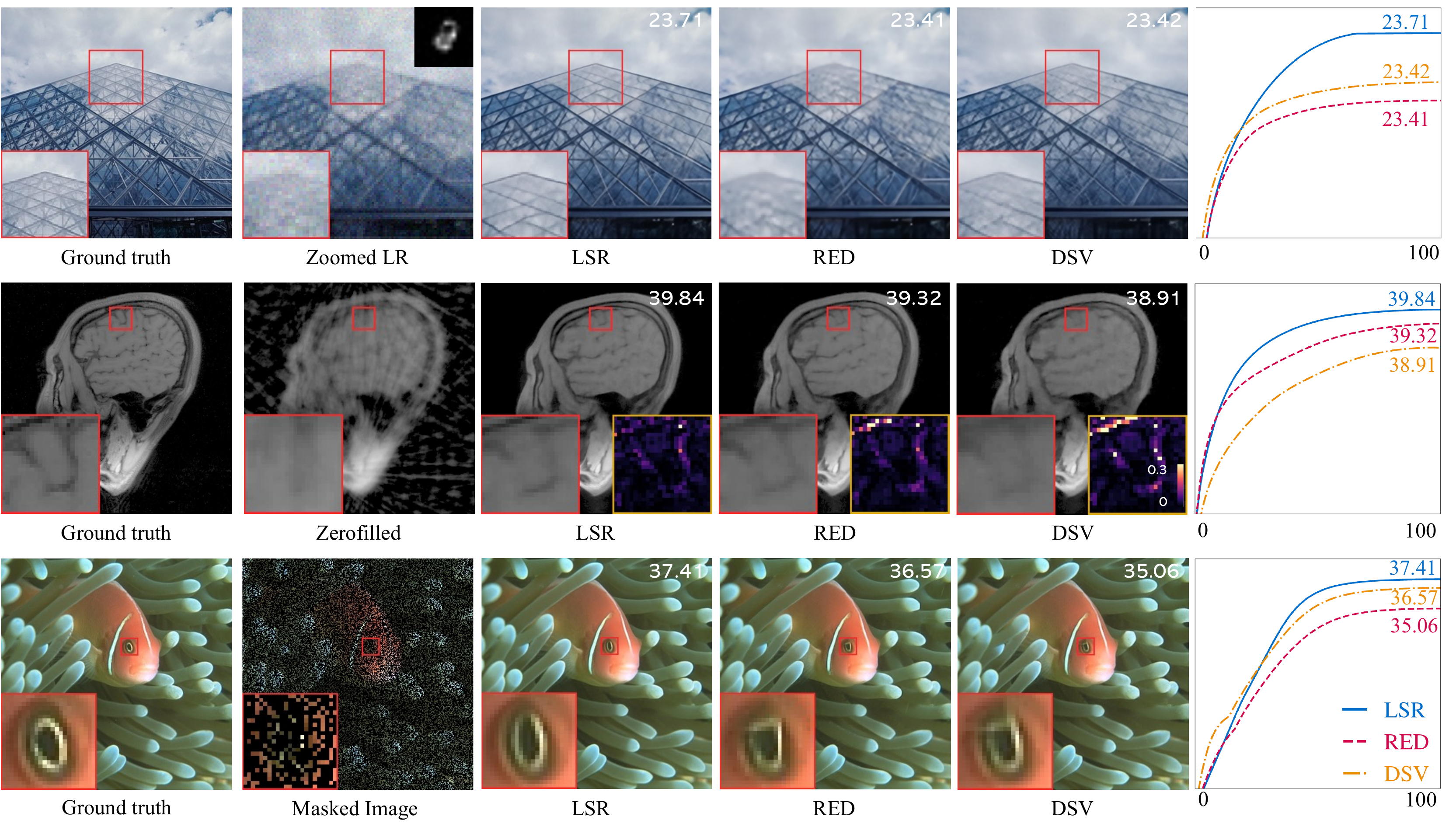}
	\caption{Visual illustration of the results obtained by ELDER using the three parameterization approaches for $h_\thetabm$. We show results for three inverse problems: image super-resolution (top), reconstruction from Fourier samples (middle), and image inpainting (bottom). The rightmost panel plots the evolution of PSNR (dB) accross forward iterations. Note how LSR achieves an overall better performance compared to DSV and RED.}
\label{fig:potential}
\end{figure*}

Table~\ref{Tab:potential_compare} and Fig.~\ref{fig:potential} present the results of comparing the three parameterization strategies within ELDER on the three inverse problems: single image super-resolution, reconstruction from Fourier measurements, and image inpainting (each problem is discussed in the dedicated section below). We observe that LSR leads to the best results, motivating its use as a primary parameterization strategy for ELDER. 
\begin{table*}[h!]
\centering
\caption{Average PSNR (dB) results of different methods on CBSD68. The \textbf{best} and \underline{second} two results are highlighted, respectively.}
\resizebox{\textwidth}{!}{
\begin{tabular}{lcccccccccccc}
\hline
 \multirow{2}{*}{\adjustbox{padding=0 0 0 2mm,valign=T}{Method}} &
   \multirow{2}{*}{\adjustbox{padding=0 0 0 2mm,valign=T}{Scale \& Noise}} &
  \multicolumn{10}{c}{Blur Kernel} &
  \multirow{2}{*}{\adjustbox{padding=0 0 0 2mm,valign=T}{Average}} \\ \cline{3-12}
 &
   &
\adjustbox{padding*=1mm,valign=M}{\includegraphics[width=7mm]{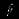}} &
  \adjustbox{padding*=1mm,valign=M}{\includegraphics[width=7mm]{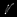}} &
  \adjustbox{padding*=1mm,valign=M}{\includegraphics[width=7mm]{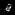}} &
  \adjustbox{padding*=1mm,valign=M}{\includegraphics[width=7mm]{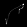}} &
  \adjustbox{padding*=1mm,valign=M}{\includegraphics[width=7mm]{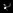}} &
  \adjustbox{padding*=1mm,valign=M}{\includegraphics[width=7mm]{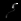}} &
  \adjustbox{padding*=1mm,valign=M}{\includegraphics[width=7mm]{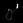}} &
  \adjustbox{padding*=1mm,valign=M}{\includegraphics[width=7mm]{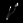}} &
  \adjustbox{padding*=1mm,valign=M}{\includegraphics[width=7mm]{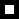}} &
  \adjustbox{padding*=1mm,valign=M}{\includegraphics[width=7mm]{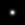}}
   &
   \\ \hline\hline
 Bicubic &
  \multirow{7}{5em}{$\sigma = 7.65$,\\ $d = 3$} &
  21.71 &
  21.96 &
  21.31 &
  19.11 &
  22.54 &
  19.36 &
  20.24 &
  20.30 &
  22.11 &
  23.06 &
  21.17 \\
 RCAN~\cite{zhang2018image} &
   &
  20.67 &
  20.99 &
  20.17 &
  18.48 &
  21.32 &
  18.44 &
  19.46 &
  19.45 &
  21.13 &
  22.14 &
  20.23 \\
 IRCNN~\cite{Zhang.etal2017a} &
   &
  24.34 &
  24.36 &
  24.98 &
  24.13 &
  25.04 &
  24.60 &
  24.49 &
  24.51 &
  24.11 &
  24.99 &
  24.56 \\
 DPIR~\cite{Zhang.etal2021b} &
   &
  24.50 &
  24.63 &
  25.51 &
  24.22 &
  25.75 &
  24.99 &
  24.92 &
  24.97 &
  24.38 &
  25.86 &
  24.97 \\
 GSPnP~\cite{Hurault.etal2022} &
   &
  24.62 &
  24.86 &
  25.63 &
  23.99 &
  25.72 &
  24.17 &
  24.55 &
  24.93 &
  24.36 &
  25.92 &
  24.88 \\
 DEQ~\cite{Gilton.etal2021} &
   &
  \underline{24.90} &
  \underline{25.03} &
  \underline{25.84} &
  \underline{24.59} &
  \underline{25.91} &
  \underline{24.69} &
  \underline{25.06} &
  \underline{25.15} &
  \underline{24.70} &
  \underline{26.14} &
  \underline{25.20} \\
 ELDER (Ours) &
   &
  \textbf{25.15} &
  \textbf{25.31} &
  \textbf{25.97} &
  \textbf{25.01} &
  \textbf{26.11} &
  \textbf{25.65} &
  \textbf{25.53} &
  \textbf{25.49} &
  \textbf{24.84} &
  \textbf{26.24} &
  \textbf{25.53} \\ \hline
 Bicubic &
   \multirow{7}{5em}{$\sigma = 2.55$,\\ $d = 4$} &
  21.60 &
  22.04 &
  21.12 &
  19.40 &
  22.31 &
  19.70 &
  20.46 &
  20.71 &
  22.28 &
  22.64 &
  21.23 \\
 RCAN~\cite{zhang2018image} &
   &
  20.84 &
  21.50 &
  20.04 &
  18.98 &
  21.32 &
  18.77 &
  19.91 &
  20.11 &
  21.84 &
  21.91 &
  20.52 \\
 IRCNN~\cite{Zhang.etal2017a} &
   &
  24.17 &
  24.23 &
  24.97 &
  24.05 &
  24.89 &
  24.46 &
  24.12 &
  24.36 &
  24.43 &
  25.18 &
  24.49 \\
 DPIR~\cite{Zhang.etal2021b} &
   &
  24.35 &
  24.58 &
  25.33 &
  24.14 &
  25.21 &
  24.78 &
  24.29 &
  24.80 &
  24.79 &
  25.57 &
  24.78 \\
 GSPnP~\cite{Hurault.etal2022} &
   &
  24.03 &
  24.25 &
  25.09 &
  22.88 &
  24.89 &
  23.55 &
  23.21 &
  23.88 &
  24.72 &
  25.41 &
  24.19 \\
 DEQ~\cite{Gilton.etal2021} &
   &
  \underline{24.85} &
  \underline{24.86} &
  \underline{25.62} &
  \underline{24.93} &
  \textbf{25.53} &
  \textbf{25.37} &
  \textbf{25.11} &
  \textbf{25.26} &
  \textbf{25.17} &
  \textbf{25.87} &
  \textbf{25.26} \\
 ELDER (Ours) &
   &
  \textbf{24.85} &
  \textbf{25.00} &
  \textbf{25.67} &
  \textbf{24.96} &
  \underline{25.45} &
  \underline{25.29} &
  \underline{25.05} &
  \underline{25.18} &
  \underline{25.07} &
  \underline{25.74} &
  \underline{25.22} \\ \hline
\end{tabular}%
}

\label{Tab:SuperResolution}
\end{table*}

\subsection{Single Image Super-Resolution.}
\begin{figure*}[t!]
	\centering
	\includegraphics[width=0.99\linewidth]{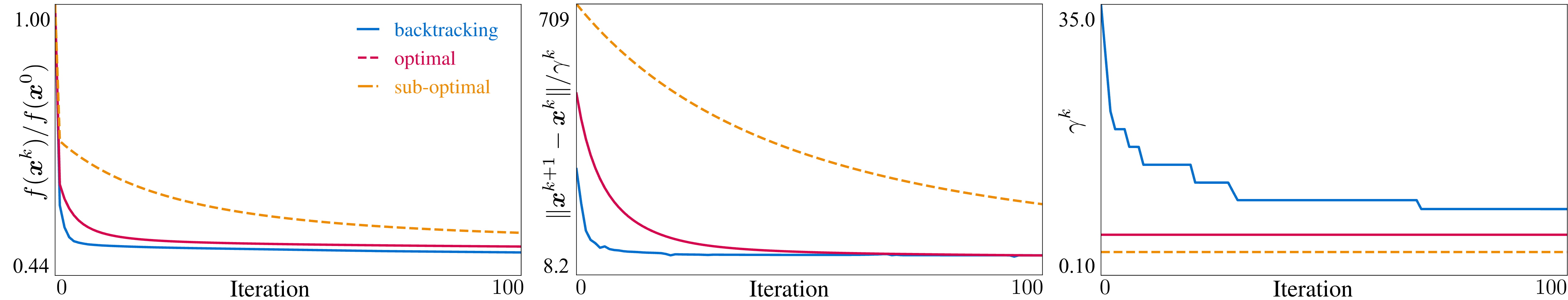}
	\caption{Illustration of the convergence of the ELDER forward iterations on SISR using three step-size selection strategies: (a) backtracking, (b) optimal, and (c) sub-optimal. Optimal is the idealized strategy that uses MSE with respect to the ground truth for step-size selection. Backtracking is a practical strategy that uses a backtracking line-search for automatic step-size selection. Suboptimal refers to a step-size selected as $10\%$ of the optimal one. \emph{Left}: Convergence of the objective. \emph{Middle}: Convergence of the updates. \emph{Right}: Step-size values across iterations. This figure illustrates that one of the key benefits of ELDER is that it can use backtracking for automatic step-size selection.}
	\label{Fig:Converge}
\end{figure*} 

\begin{figure*}[t!]
	\centering
	\includegraphics[width=0.99\linewidth]{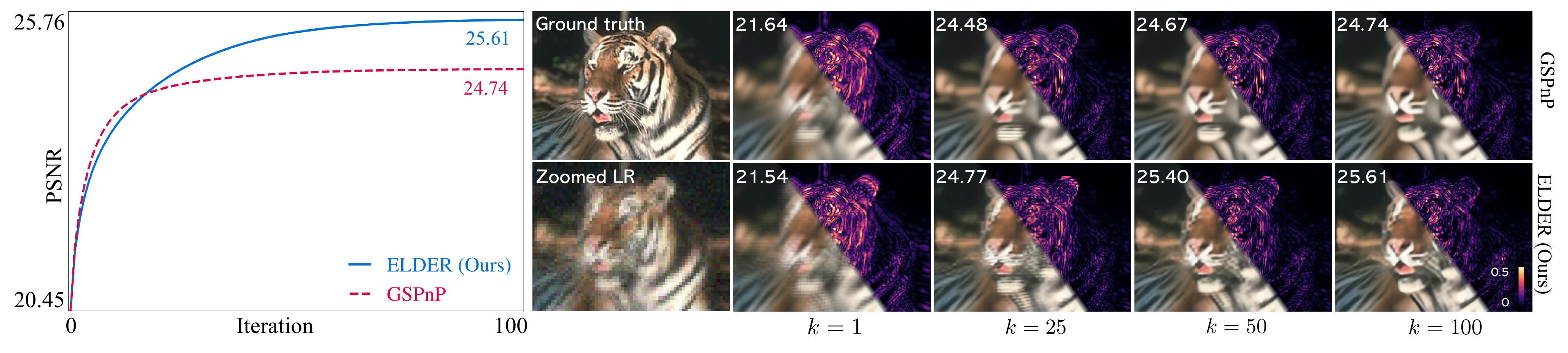}
	\caption{Comparison between ELDER and GSPnP for SISR with $d=3$. Both methods seek to learn explicit regularizers parametrized by the same CNN architecture, but are trained differently. \emph{Left}: Evolution of the PSNR is plotted against iteration number for both methods. \emph{Right}: Visual illustration of iterates obtained by GSPnP (top) and ELDER (bottom). Note how ELDER outperforms GSPnP due to its ability to learn regularizers that minimize MSE.}
	\label{Fig:PSNR}
\end{figure*} 
We consider measurements of form $\ybm = \Sbf\Hbf\xbm + \ebm$, where $\Hbf\in\R^{n\times n}$ is the blurring matrix and $\Sbf\in\R^{m\times n}$ performs standard $d$-fold down-sampling with $d^2=n/m$. When the blur satisfies the circular boundary conditions, the blurring matrix and its conjugate transpose can be decomposed as $\Hbf=\Fbf^{\Hsf}\Mbf\Fbf$ and $\Hbf^{\Hsf}=\Fbf^{\Hsf}\Mbf^{\Hsf}\Fbf$, where $\Fbf$ is the discrete Fourier transform (and $\Fbf^{\Hsf}$ its inverse, satisfying $\Fbf\Fbf^{\Hsf}=\Fbf^{\Hsf}\Fbf=\Ibf$), and $\Mbf\in\C^{n\times n}$ is a diagonal matrix whose diagonal elements are the Fourier coefficients of the blur. The proximal operator~\eqref{Eq:Prox_GS-PnP} of $g(\xbm) = \frac{1}{2}\|\ybm - \Sbf\Hbf\xbm\|_2^2$ has the closed-form solution
\begin{equation}
\label{Eq:srprox}
\prox_{\gamma g} (\zbm) = \rbm - \Fbf^{\Hsf}\left(\frac{{{\bf\underline{\Mbf}}^{\Hsf}\bf\underline{\Mbf}}\Fbf\rbm}{d^2\Ibf + \gamma{\bf\underline{\Mbf\Mbf}^{\Hsf}}}\right),
\end{equation}
where $\rbm=\gamma\Hbf^{\Hsf}\Sbf^{\Hsf}\ybm + \zbm$. The matrix ${\bf\underline{\Mbf}} = [{\bf{\Mbf}}_1,\cdots,{\bf{\Mbf}}_{d^2}]\in\C^{m\times n}$, and where the blocks ${\bf{\Mbf}}_{i}\in\C^{m\times n}$ satisfy ${\bf{\Mbf}}=\textbf{diag}\{{\bf{\Mbf}}_{1},\cdots,{\bf{\Mbf}}_{d^2}\}$ a block-diagonal decomposition according to a $d \times d$ tiling in the Fourier domain (see Lemma 1 in~\cite{Zhao.etal2016}). Note that the inverse of diagonal matrix $d^2\Ibf + {\bf\underline{\Mbf\Mbf}^{\Hsf}}$ can be computed element-wise. 

We verify the effectiveness of ELDER on SISR using a large variety of blur kernels and different down-sampling factors. We use $8$ realistic camera shake kernels tested in~\cite{Zhang.etal2021b}, plus a $9\times 9$ uniform kernel, and a $25\times 25$ Gaussian kernel with standard deviation $1.6$. All $10$ kernels are presented in Table~\ref{Tab:SuperResolution}. We use the same dataset in Section~\ref{Sec:Imagedenoising} for training ELDER. We train a single model on all blur kernels, each with downsampling factors of $d\in\{2,3,4\}$, to test the generalizability under different SISR settings. We set the number of forward-iterations to $K=100$. At every training iteration, we initialize $\xbm_0$ with a shift-corrected bicubic interpolation of $\ybm$~\cite{Zhang.etal2021b}. Additionally, during training we use AWGN with random noise levels $\sigma\in[0,10.0]/255$ to ensure robustness of our model to different noise perturbations.
\begin{figure*}[t!]
	\centering
	\includegraphics[width=\textwidth]{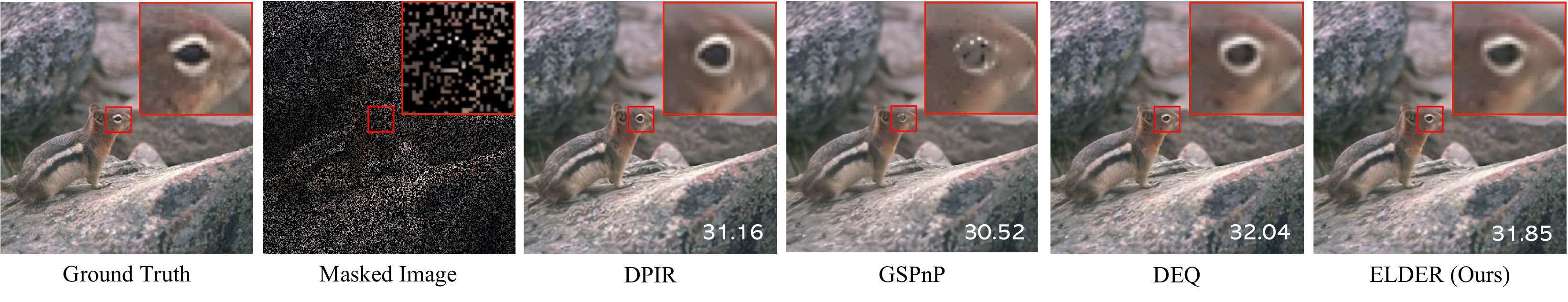}
	\caption{Visual illustration of ELDER relative to DPIR, GSPnP and DEQ for image inpainting with masked probability of $p=0.7$ on CBSD68. Note that ELDER provides substantial improvements over PnP image reconstruction methods, matching the performance of DEQ.
	}
	\label{Fig:Inpainting}
\end{figure*}
\begin{figure*}[t!]
	\centering
	\includegraphics[width=\textwidth]{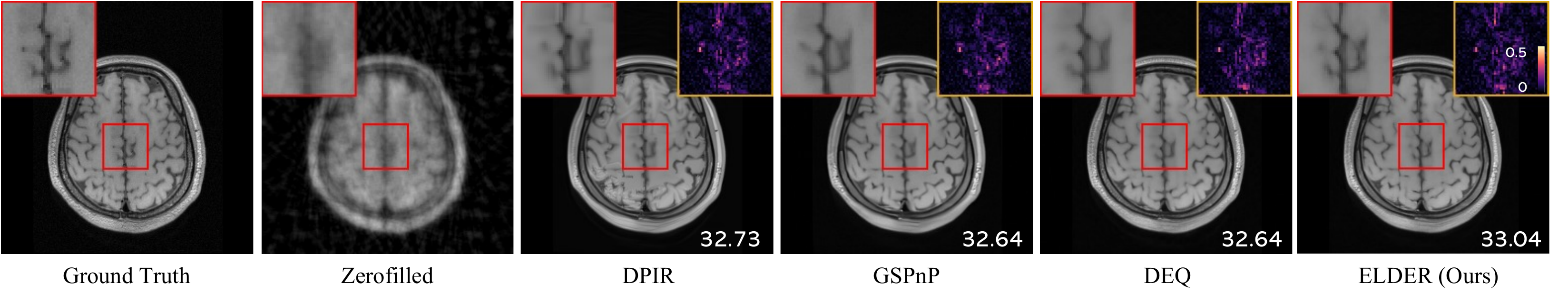}
	\caption{Visual evaluation of several well-known methods on reconstruction of a brain image from its radial Fourier measurements at $10\%$ sampling. Note how ELDER outperforms DPIR, GSPnP, and the traditional DEQ both quantitatively and visually.
	}
	\label{Fig:CSMRI}
\end{figure*}

We compare ELDER and DEQ against bicubic interpolation, RCAN~\cite{zhang2018image}, and state-of-the-art PnP methods IRCNN~\cite{Zhang.etal2017a}, DPIR~\cite{Zhang.etal2021b}, and GSPnP~\cite{Hurault.etal2022}. We use the publicly available implementations for all the baseline methods. RCAN refers to the PSNR oriented deep model based on bicubic degradation. GSPnP~\cite{Hurault.etal2022} is a PnP method using pre-trained denoising CNN as an explicit regularization functional, with an algorithmic update similar to the ELDER forward pass. DPIR uses the original DRUNet in Table~\ref{Tab:Denoising}, while GSPnP uses the LSR gradient-step denoiser. The regularization and step-size parameters of GSPnP, DEQ, and ELDER were optimized at test time using \texttt{fminbound} in \texttt{scipy.optimize}.

\begin{table*}[t]
\centering
\caption{Average PSNR (dB) and SSIM values for different methods on CS-MRI.}
\resizebox{\linewidth}{!}{
\begin{tabular}{p{45pt}C{45pt}C{45pt}C{70pt}C{45pt}C{45pt}C{45pt}C{45pt}C{45pt}C{45pt}}
\hline
 {  } & \multicolumn{8}{c}{ {Method}} \\ \cline{3-10}
\multirow{-2}{*}{CS Ratio}     & \multirow{-2}{*}{Metric}                   & TV~\cite{Beck.Teboulle2009}   & ADMM-Net~\cite{Yang.etal2016} & IRCNN~\cite{Zhang.etal2017a} & DPIR~\cite{zhang.etal2021} & ISP~\cite{Cohen.etal2021} & GSPnP~\cite{Hurault.etal2022} & DEQ~\cite{Gilton.etal2021} &ELDER\\ \hline\hline

\multirow{2}{*}{10\%}          & PSNR      & 31.36 & 34.19 & 33.86 & 34.98 & 34.27 & 34.86  & \textbf{35.16} & \underline{35.14}            \\ 
                               & SSIM      & 0.8200 & 0.8959 & 0.8865 & 0.9060 & 0.9007 & 0.9063 & \underline{0.9114} & \textbf{0.9116}      \\
\multirow{2}{*}{20\%}          & PSNR      & 35.62 & 37.17 & 37.84 & 38.70 & 38.35 & 38.56 & \textbf{38.88} & \underline{38.68}              \\ 
                               & SSIM      & 0.9121 & 0.9471 & 0.9411 & 0.9464 & 0.9414 & 0.9478 &  \textbf{0.9499} & \underline{0.9492}       \\\cdashline{1-10}
\multirow{2}{*}{Average}       & PSNR      & 33.49 & 35.68 & 35.52 & 36.84 & 36.31 & 36.71 & \bf{37.02} & \underline{36.91}                \\ 
                               & SSIM      & 0.8660 & 0.9215 & 0.9138 & 0.9262 & 0.9211 & 0.9271 & \bf{0.9307} & \underline{0.9304}         \\\hline 
\end{tabular}%
}
\label{Tab:CSMRI}
\end{table*}
Table~\ref{Tab:SuperResolution} summarizes the PSNR values achieved by ELDER and other methods when applied to $\{\sigma=7.65, d=3\}$ and $\{\sigma=2.55, d=4\}$ on the CBSD68 dataset. It is clear that both ELDER and DEQ outperform the other methods at all settings. Additionally, ELDER matches the overall performance of DEQ, while also performing better for higher noise levels at $d = 3$. The excellent performance of ELDER relative to the traditional DEQ highlights that use of an explicit regularizer does not imply a compromise in performance.
Fig.~\ref{fig:sr_visual} visually compares ELDER against the baseline methods at scale factors $\times 3$ (top) and $\times 4$ (bottom), respectively. The enlarged regions in the images suggest that ELDER better recovers the fine details and sharper edges compared to DPIR and GSPnP, while providing same or better PSNR than DEQ. These results indicate that ELDER reaches state-of-the-art performance in PnP/DEQ SISR for a variety of kernels and noise levels. 

Fig.~\ref{Fig:Converge} illustrate the convergence behavior of the forward pass of ELDER in terms of the objective $f(\xbm^k)$ (left) and the residual $\|\xbm^{k}-\xbm^{k-1}\|_2/\gamma$ (middle), when tested on a subset of $10$ color images taken from the original CBSD68 dataset (CBSD10). Fig.~\ref{Fig:Converge} (right) shows the corresponding step-size $\gamma$ used in the experiment. The optimal step-size on each test image is optimized for the best PSNR value. Note how ELDER using the backtracking line-search strategy enables automatic selection of the step-size parameter. 
\begin{table}[t]
\centering
\caption{Average PSNR (dB)/SSIM values for different methods for image inpainting on CBSD68 dataset.}
\label{Tab:Inpainting}
\resizebox{0.6\linewidth}{!}{
\begin{tabular}{lccc}
\hline
 {  } & \multicolumn{2}{c}{ {Probability of Masking}}          &  {  } \\ \cline{2-3}
\multirow{-2}{*}{Method}                        & 50\%   & 70\%       &   \multirow{-2}{*}{Average}                       \\ \hline\hline
IRCNN                   & 31.61/0.9227 & 27.87/0.8558   & 29.77/0.8893                   \\ 
DPIR                    & 31.72/0.9274 & 28.11/0.8601   & 29.92/0.8938                   \\ 
GSPnP                   & 31.66/0.9263 & 27.98/0.8251  & 29.82/0.8757                   \\ 
DEQ                     & \bf32.79/0.9432 & \bf29.31/0.8820  & \bf31.05/0.9126                   \\ 
ELDER (Ours)               & \underline{32.30/0.9352} & \underline{28.82/0.8700}  & \underline{30.56/0.9026}                   \\ \hline
\end{tabular}%
}
\end{table}

Fig.~\ref{Fig:PSNR} compares the convergence of ELDER and GSPnP in terms of PSNR for SISR with scale factor 3. The figure also shows visual results for both methods at different iterations. Note how ELDER learns a regularizer that leads to an improved PSNR compared to GSPnP. Since both methods share the same parametrized regularization functional, the improvement is due to DEQ learning of the regularizer.

\subsection{Compressed Sensing MRI (CS-MRI)}
MRI is a widely-used medical imaging technology that is limited by the low speed of data acquisition. CS-MRI seeks to address this limitation by recovering an image $\xbmast$ from its sparsely-sampled Fourier measurements. We simulate a simplified noiseless single-coil CS-MRI using radial Fourier sampling. The measurement operator is thus $\Abf=\Bbf\Fbf$, where $\Bbf$ is the diagonal sampling matrix with values in $\{0,1\}$. The proximal operator of $g(\xbm)=\frac{1}{2}\|\ybm - \Bbf\Fbf\xbm\|_2^2$ has a closed-form
\begin{equation}
\label{Eq:mriprox}
\prox_{\gamma g} (\zbm) =\Fbf^{\Hsf}\left(\frac{{\gamma{\bf{\Bbf}}^{\Hsf}}\ybm + \Fbf\zbm}{\Ibf + \gamma{\bf{\Bbf}^{\Hsf}\Bbf}}\right).
\end{equation}

\noindent
We train ELDER using the brain dataset from~\cite{zhang2018ista}, which consists of 800 slices of $256 \times 256$ training images and 50 slices of testing images. Both ELDER and conventional DEQ are trained on sampling ratios within $\left[10,20\%\right]$. At every training iteration, we use zero-filled image $\xbm_0=\Abf^{\Hsf}\ybm$ to initialize the forward pass with $100$ iterations. 

Table \ref{Tab:CSMRI} presents average PSNR values obtained by ELDER, publicly available implementations of several well-known methods, including TV~\cite{Beck.Teboulle2009}, ADMM-Net~\cite{Yang.etal2016}, as well as IRCNN, DPIR, GSPnP, ISP~\cite{Cohen.etal2021}, and DEQ. Specifically, TV is solved using the accelerated proximal gradient descent method~\cite{Beck.Teboulle2009}. ADMM-Net is a deep unrolling method that trains both image transforms and shrinkage functions within the algorithm. ISP refers to an MBDL using explicit deep denoisers, similar to GSPnP. Overall, ELDER and DEQ achieve the best performances, indicating that having the explicit regularizer does not compromise performance. Fig.~\ref{Fig:CSMRI} provides some visual examples at sampling ratio $10\%$, highlighting the imaging quality obtained by our method relative to several baseline methods. From the zoomed regions and the corresponding error maps, ELDER improves over DPIR and GSPnP due to the training of the explicit regularizer to be end-to-end  MSE optimal. 

\subsection{Image Inpainting}

We now apply ELDER to image inpainting characterized by the measurement model $\ybm=\Pbf\xbm$, where $\Abf=\Pbf$ is a random diagonal matrix with $p \in [0,1]$ denoting a probability of missing a pixel. We assume a noiseless setting. In this context, the data-fidelity term is the indicator function $g(\xbm) =\Pi_{\Ccal}(\xbm)$ for the set $\Ccal=\{\xbm\in\R^n \,:\, \ybm=\Pbf\xbm\}$, which, by definition is $0$ when $\xbm \in \Ccal$ and $+\infty$ elsewhere. The proximal step~\eqref{Eq:Prox_GS-PnP} has a closed form solution
\begin{equation}
\label{Eq:inpaintprox}
\prox_{\gamma g} (\zbm) = \ybm + \zbm - \Pbf\zbm.
\end{equation}
At the $k$th forward iteration of  ELDER, the proximal operator returns an image consisting of the network output at the missing pixels and measured pixels at the other locations.

We train our model under  random  sampling parameter $p\in[0.3,0.7]$. To demonstrate the flexibility of ELDER, we consider $p=0.5$ and $p=0.7$ and compare to IRCNN, DPIR, GSPnP, and DEQ. Again, since IRCNN and DPIR lack the implementation for inpainting, we apply our data fidelity term on them, set the $\sigma_K$ parameter to 3.0, and set the iteration number to 100. The PSNR performance is reported in Table~\ref{Tab:Inpainting}, and visual results are provided in Fig.~\ref{Tab:Inpainting}. These results  indicate that, achieves better performance compared to GSPnP and nearly matches the performance of DEQ. Moreover, we can observe from the visual results that GSPnP smoothes out the fine details, while DPIR generates distortions. In contrast, ELDER can recover detail as well as avoid distortions.

\begin{table}[t]
\centering
\caption{Average PSNR (dB) values for different explicit CNN regularizer used in ELDER on SISR, CS-MRI and Inpainting.}
\label{Tab:potential_compare}
\resizebox{0.6\columnwidth}{!}{%
\begin{tabular}{lccccccc}
\hline
\multirow{2}{*}{Regularizers} &
                              & \multicolumn{2}{c}{SISR} & \multicolumn{2}{c}{CS-MRI} & \multicolumn{2}{c}{Inpainting} \\ \cline{3-8} 
                              && x3  & x4 &          10\%     &20\%          & 50\%           & 70\%          \\ \hline\hline
LSR                        && 25.53       & 25.22      & 35.14       & 38.68            & 32.30          & 28.82          \\
RED                       && 25.42       & 25.17      & 34.76       & 38.21            & 32.22          & 28.55         \\
DSV                       && 25.42       & 25,19      & 34.07       & 37.87            & 32.14          & 28.77         \\ \hline
\end{tabular}%
}
\end{table}

\section{Conclusion}

We present ELDER as a novel framework for learning explicit regularizers for model-based deep learning in imaging inverse problems. ELDER parameterizes the regularizer using a CNN and learns its weights to minimize MSE values using DEQ. The key benefit of having an explicit regularizer is that one directly inherits the fundamental results from optimization theory. We show that ELDER outperforms existing approaches for learning explicit regularizers for inverse problems. It is also worth noting that this work suggests that using an \emph{explicit} regularization functional does \emph{not} compromise imaging performance compared to methods based on \emph{implicit} regularization. 

\section*{Acknowledgements}

This material is based upon work supported by the NSF CAREER award under grant CCF-2043134, and by the Laboratory Directed Research and Development program of Los Alamos National Laboratory under project number 20200061DR.

\bibliographystyle{IEEEtran}


\end{document}